\documentclass{revtex4}
\usepackage{amsmath}

\begin{document}

\title{Gravitational trapping potential with arbitrary extra dimensions}

\author{Douglas Singleton}
\email{dougs@csufresno.edu}
\affiliation{Centre of Gravitation and Fundamental Metrology, VNIIMS, 3-1 M.
Ulyanovoy St., Moscow 119313, Russia; \\
Physics Dept., CSU Fresno, 2345 East San Ramon Ave.
M/S 37 Fresno, CA 93740-8031, USA}

\date{\today}

\begin{abstract}
We extend a recently discovered, non-singular 6 dimensional brane,
solution to $D=4+n$ dimensions. As with the previous 6D solution
the present solution provides a gravitational trapping mechanism
for fields of spin $0, \frac{1}{2}, 1$ and $2$. There is an
important distinction between 2 extra dimensions and $n$ extra
dimensions that makes this more than a trivial extension. In contrast
to gravity in $n >2$ dimensions, gravity in $n=2$ dimensions is
conformally flat.  The stress-energy tensor required by this
solution has reasonable physically properties, and for $n=2$ and
$n=3$ can be made to asymptotically go to zero as one moves away
from the brane.
\end{abstract}


\maketitle

\section{Introduction}

The brane world models introduced in \cite{Go, RaSu} have provided
a new framework for addressing various questions in particle
physics, such as the hierarchy problem, using non-compact extra
dimensions (see \cite{Ak, Ru, Vi, Gib, Mel1, Mel2} for earlier work on non-compact
extra dimensions and on solutions in extra dimensions). In these models one must give a 
mechanism for trapping or localizing fields with various spins ($0, \frac{1}{2},
1, 2$) on the 4 dimensional brane of the observable Universe. The
condition for determining whether a multi-dimensional field is
localized or not is if the integral of its Lagrangian over the
extra coordinates is finite. In the original 5 dimensional models
it was found that it was not possible to gravitational trap all
types of fields with the same warp factor. Spin $0$ and spin $2$
fields were trapped on the brane with a decreasing, exponential,
warp factor, while spin $\frac{1}{2}$ fields were trapped with an
increasing warp factor \cite{RaSu, BaGa}. Spin $1$ fields were not
localized at all \cite{Po}.  In the 6 dimensional case with two
extra dimensions slightly better trapping behavior was found
\cite{Od}: $0$, $1$ and $2$ fields were localized with a
decreasing, exponential warp factor but spin $\frac{1}{2}$ fields
were again localized with an increasing, exponential warp factor.
In \cite{GoMi} it was shown that one could localize the zero modes
of all these fields with an increasing, non-exponential warp
factor in 6 dimensions with one extra time-like dimension. In
\cite{GoSi, GoSi1} it was shown that all these fields could be
localized in 6 dimensions with both extra dimensions being
space-like. In particular \cite{GoSi, Mi} gave an exact,
non-singular, non-exponential trapping solution. Given the
improving localization behavior with increasing spacetime
dimension the natural inclination would be to look for trapping
solutions with still more extra dimensions. However, it was pointed
out in \cite{GhRo} that there is an important distinction in going
from $n=2$ to $n >2$ extra dimensions. Gravity in 2 dimensions is
conformally flat. For spacetime with greater than 2 dimensions
vacuum spacetime can have a non-zero Weyl tensor. The present work shows
that it is possible to extend (with some restrictions) the non-singular, trapping solution
of \cite{GoSi, Mi} to $D=4+n$ dimensions. (Earlier work on the localization 
of gravity in $n \ge 2$ extra dimensions with higher curvature terms
can be found in \cite{Neu}). The solution localizes
all types of spin fields with a non-exponential, warp factor. The
draw back of the solution is that the stress-energy tensor is
selected or tuned so as to give these good features of the
solutions. However the stress-energy tensor is physically
reasonable in that its magnitude is peaked on the brane and
decreases as one moves away from the brane. It will be shown that
for the $n=2$ and $n=3$ cases it is possible to have the
stress-energy tensor go to zero as one moves away from the brane.
For $n\ge 4$ one can have part, but not all of the stress-energy
tensor asymptotically go to zero. 

It is possible to localize all spin fields in the 5D and 6D models
with exponential warp factors if one introduces non-gravitational
interactions. However, for reasons of simplicity, and since
gravity couples universally to all types of matter, it is
preferable to find a purely gravitational trapping mechanism. The
solution presented here does give such a pure gravitational
trapping mechanism for all types of spin fields. The warp factor
is non-exponential in contrast to the work in \cite{Go,RaSu,GhSh,
Gr}

\section{D-dimensional Einstein Equations}

The system we consider is $D=4+n$ dimensional gravity with a
cosmological constant $\Lambda$, and some matter fields. The
action is
\begin{equation} \label{action}
S = \int d^D x\sqrt{- ^D g}\left[\frac{M^{n+2}}{2}(^D R + 2
\Lambda) + L_m \right] ~,
\end{equation}
where $\sqrt{-^D g}$ is the determinant, $M$ is the fundamental
mass scale, $^D R$ is the scalar curvature, $\Lambda$ is the
cosmological constant and $L$ is the Lagrangian of matter fields.
All of these quantities are $D=4+n$ dimensional. The
D-dimensional Einstein equations resulting from this action are
\begin{equation}
\label{Einstein-n}
    ^D R_{AB} - \frac{1}{2} g_{AB} ~^D R = \Lambda
   g_{AB} + \frac{1}{M^{n+2}} T_{AB}~,
\end{equation}
Capital Latin indices run over $A, B,... = 0, 1, 2, 3, ..., D$.
The stress-energy tensor, $T_{AB}$, is determined from the matter
Lagrangian, $L_m$, in the standard way.

The ansatz we will use is a simple extension of refs. \cite{GoSi, GoSi1}
with two scale functions -- one for the 4-dimensional brane and one
for the $n$ extra spatial dimensions.
\begin{equation}
\label{ansatz}
ds^2= \phi ^2(r) \eta_{\alpha \beta }(x^\nu)dx^\alpha dx^\beta -
\lambda (r) (dr^2 +  r^2 d \Omega ^2 _{n-1}) ~.
\end{equation}
The metric of ordinary 4-dimensional spacetime, $\eta_{\alpha
\beta }(x^\nu)$, has the signature $(+,-,-,-)$. The Greek indices
$\alpha, \beta,... = 0, 1, 2, 3$ refer to the coordinates of these
4-dimensions. It is assumed that ansatz \eqref{ansatz} only
depends on the extra coordinate $r$ through the 4-dimensional
conformal factors, $\phi^2$, and $\lambda$. $d \Omega ^2 _{n-1}$
is the solid angle for the $(n-1)$ sphere and is explicitly given
by
\begin{equation}
\label{solidangle} d \Omega ^2 _{n-1} = d \theta ^2 _{n-1} + \sin
^2 \theta _{n-1} d \theta _{n-2} ^2 + \sin ^2 \theta _{n-1} \sin
^2 \theta _{n-2} d \theta _{n-3} ^2 + \cdots + \prod _{i=2} ^{n-1}
\sin ^2 \theta _i d \theta _1 ^2
\end{equation}
The angles have the following ranges: $0 \le \theta _1 \le 2 \pi$
and $0 \le \theta _i \le \pi$ for $i=2, ..., n-1$. The radial
coordinate has the range $0 \le r < \infty$.

The above generalization of the 6D ansatz of \cite{GoSi, GoSi1}
has an important difference from the generalization of the 6D
metric ansatz considered in \cite{Od,GhSh,Gr}. In these latter papers the scale
factor, $\lambda (r)$,  multiples only the angular part of the metric
\begin{equation}
\label{gher}
ds^2  = \phi ^2 (r)\eta _{\alpha \beta } (x^\nu
)dx^\alpha dx^\beta   - dr^2  - \lambda (r)d\theta ^2~,
\end{equation}
whereas in \cite{GoSi, GoSi1} the scale factor multiplies the
entire extra dimensional part.  Both the 6D version of  \eqref{ansatz} and
\eqref{gher} are locally the same. This can be seen by defining $dr$ of
\eqref{ansatz} via $\lambda (r) dr^2 = d{\tilde r} ^2$ where 
${\tilde r}$ is the radial coordinate of \eqref{gher}. However there is a global
distinction between \eqref{ansatz} and \eqref{gher}. The metric
in \eqref{gher} corresponds to a cone-like geometry such as occurs with 
string-like defect. For example, if in \eqref{gher} one has 
$\lambda (r) = (1 - n)^2 r^2$ then the geometry of the extra space will
exhibit an angular deficit of $2 \pi n$. A similar $\lambda (r)$ in \eqref{ansatz}
would not result in an angular deficit.

For the 6D ansatz of \cite{GoSi, GoSi1} and its
generalization to 4+n given in \eqref{ansatz} the metric
functions, $\lambda (r)$ and $\phi (r)$, are conformal factors for
an n-dimensional Euclidean metric and 4-dimensional Minkowski
metric respectively. Thus as long as the metric functions are well
behaved ({\it i.e.} not equal to $0$ or $\infty$) the metric will
be non-singular. Since we are looking for non-singular solutions,
we choose the metric to take the form \eqref{ansatz}.

The nonzero components of the stress-energy tensor $T _{AB}$ are
taken to be of the form
\begin{equation} \label{source}
T_{\mu\nu} = - g_{\mu\nu} F(r), ~~~ T_{ij} = - g_{ij}K(r),
     ~~~ T_{i\mu} = 0 ~.
\end{equation}
The two source functions, $F(r)$ and $K(r)$, depend only on the
radial coordinate $r$.

As a final simplification we assume that 4-dimensional metric is
flat ($\eta _{\alpha \beta} (x^\nu) = \eta _{\alpha \beta}$) and
that the 4-dimensional cosmological constant is zero. This yields
\begin{equation} \label{Einstein4}
^4 R_{\mu\nu} - \frac{1}{2} \eta_{\mu\nu} {^4R} = 0~,
\end{equation}
where $^4R_{\alpha\beta}$, $^4R$ are the 4-dimensional Ricci
tensor and scalar constructed from $\eta_{\alpha\beta}$. With
these simplifications and the ans{\"a}tze of \eqref{ansatz} and
\eqref{source} the D-dimensional Einstein field equations
\eqref{Einstein-n} become \cite{Od1}
\begin{eqnarray}
\label{Einstein-na} 3 \left( 2\frac{\phi ^{\prime \prime}}{\phi} -
\frac{\phi ^{\prime}}{\phi} \frac{\lambda ^{\prime}}{\lambda }
\right) + 6 \frac{(\phi ^{\prime})^2}{\phi ^2} + (n-1) \left[ 3
\frac{\phi ^{\prime}}{\phi} \frac{(r^2 \lambda) ^\prime }{r^2
\lambda} +  \frac{(r^2 \lambda)^{\prime \prime}}{r^2 \lambda } +
\frac{n-4}{4} \left(\frac{(r^2 \lambda) ^\prime }{r^2 \lambda}
\right) ^2 -\frac{1}{2} \frac{\lambda
^{\prime}}{\lambda}\frac{(r^2 \lambda) ^\prime }{r^2 \lambda}
- \frac{n-2}{r^2} \right] \nonumber  \\
= 2 \lambda \left(\frac{F(r)}{M^{n+2}} - \Lambda \right) ~,  \\
\label{Einstein-nb} 12 \frac{(\phi ^{\prime})^2}{(\phi)^2 } +
(n-1) \left[ 4 \frac{\phi ^{\prime}}{\phi}\frac{(r^2 \lambda)
^\prime }{r^2 \lambda} + \frac{(n-2)}{4} \left(\frac{(r^2 \lambda)
^\prime }{r^2 \lambda} \right) ^2  - \frac{(n-2)}{r^2} \right] =
2 \lambda \left( \frac{K(r)}{M^{n+2}} -  \Lambda \right) ~, \\
\label{Einstein-nc} 4 \left( 2\frac{\phi ^{\prime \prime}}{\phi} -
\frac{\phi ^{\prime}}{\phi} \frac{\lambda ^{\prime}}{\lambda }
\right) + 12 \frac{(\phi ^{\prime})^2}{\phi ^2} + (n-2)\left[ 4
\frac{\phi ^{\prime}}{\phi} \frac{(r^2 \lambda) ^\prime }{r^2
\lambda} +  \frac{(r^2 \lambda)^{\prime \prime}}{r^2 \lambda } +
\frac{n-5}{4} \left(\frac{(r^2 \lambda) ^\prime }{r^2 \lambda}
\right) ^2 -\frac{1}{2} \frac{\lambda
^{\prime}}{\lambda}\frac{(r^2 \lambda) ^\prime }{r^2 \lambda}
- \frac{n-3}{r^2} \right] \nonumber \\
 = 2 \lambda\left(\frac{K(r)}{M^{n+2}} - \Lambda \right)~,
\end{eqnarray}
where the prime $=\partial / \partial r$. These equations are for
the $\alpha \alpha$, $rr$, and $\theta \theta$ components
respectively.

Subtracting the $rr$ component  of \eqref{Einstein-nb}  from the
$\theta \theta$ component of \eqref{Einstein-nc} to
cancel the source and cosmological terms yields
\begin{equation}
\label{phi-g} 8 \left( \frac{\phi ^{\prime \prime}}{\phi} -
\frac{\lambda  ^{\prime}}{\lambda } \frac{\phi ^{\prime}}{\phi } -
\frac{\phi ^{\prime}}{r \phi }\right) + (n-2) \left( \frac{\lambda
^{\prime \prime}}{\lambda}- \frac{3}{2}\left(\frac{\lambda
^{\prime}}{\lambda }\right)^2 - \frac{\lambda  ^{\prime}}{r
\lambda }\right) = 0 ~.
\end{equation}
For $n=2$ one has the simplification that the second term above
does not appear. Following \cite{Od1} we look for solutions in
which each term in parentheses above is zero separately
\begin{equation}
\label{g} \frac{\phi ^{\prime \prime}}{\phi} - \frac{\lambda
^{\prime}}{\lambda } \frac{\phi ^{\prime}}{\phi } - \frac{\phi
^{\prime}}{r \phi } = 0 ~~~~,~~~~  \frac{\lambda ^{\prime
\prime}}{\lambda}- \frac{3}{2}\left(\frac{\lambda
^{\prime}}{\lambda }\right)^2 - \frac{\lambda  ^{\prime}}{r
\lambda } = 0
\end{equation}
A solution to the first equation in \eqref{g} is
\begin{equation}
\label{solution1} \lambda (r) = \rho ^2 \frac{\phi ^{\prime}
(r)}{r}~,
\end{equation}
where $\rho$ is an integration constant with dimensions of length.
In \cite{Od1} it was shown that the trapping solution of \cite{GoSi1} 
could be generalized to more than two extra dimensions. The
following metric functions
\begin{equation}
\label{solution2} \phi (r) = a \frac{r^2 -c^2}{r^2 +c^2} ~~~~,~~~~
\lambda (r) = \frac{4 a c^2 \rho ^2}{(r^2 +c^2)^2}~,
\end{equation}
solve \eqref{g}, with $a$ and $c$ constants. In \cite{GoSi1} the
solution was more general in that one had $r^2 , c^2 \rightarrow
r^{b} , c^{b}$. We will find a similar selection of the exponent
when we generalize the non-singular solution of \cite{GoSi, Mi} to
$n>2$ dimensions. The solution \eqref{solution2} becomes singular
at $r=c$ and thus was taken as a solution for $r > \epsilon$ where
$\epsilon$ is the brane width, and the condition, $\epsilon \ge
c$, holds. Outside $r=\epsilon$ the scale factor for 4-dimensional
space, $\phi ^2 (r)$, asymptotically approaches the value $a^2$.
If the brane width, $\epsilon$, is small then the 4 dimensional
scale factor changes very quickly from $\phi ^2 (0) =1$ to $\phi
^2 (\infty) =a^2$. In 6D the solution in \eqref{solution2} is
supported by a stress-energy tensor whose source ansatz functions,
$F(r) , K(r)$ where proportional to $\frac{1}{\phi ^2}$. The
solution $\phi (r)$ in \eqref{solution2} was shown
\cite{GoSi1,Od1} to localize fields of all spins up to 2 on the
brane.

In \cite{GoSi, Mi} a solution was given which had similar
properties to the solution of \eqref{solution2} (non-exponential
warp factor which localized fields of all spins up to spin 2), but
which was non-singular. Here it is shown that, with some modifications
and restrictions, it is possible to
extend this solution to $n$ extra spatial dimensions. This is more
than a trivial exercise since for $n=2$ the extra space is conformally 
flat, while for $n>2$ this is not necessarily the case. Also
for both $n=2$ and $n=3$ it is possible to make a choice of parameters
({\it i.e.} $\epsilon$, $a$, $\Lambda$) such that $F(r)$ and
$K(r)$ both have the nice feature of going to zero as $r \rightarrow \infty$. 
For $n \ge 4$ one can make either $F(r)$ or $K(r)$ go to zero, but not both.

It is straightforward to check that the non-singular solution of
\cite{GoSi, Mi}
\begin{equation}
\label{non-singular} \phi (r) = \frac{c^2 + a r^2}{c^2+ r^2}
~~~~,~~~~ \lambda (r) = \frac{2 (a-1) c^2 \rho ^2}{(c^2+ r^2)^2}~,
\end{equation}
satisfies both equations from \eqref{g}; $a$ and $c$ are
constants. The solution given in \cite{Mi} was more general than
\eqref{non-singular} having the form
\begin{equation}
\label{Misoln} \phi (r) = \frac{c^b + a r^b}{c^b +
r^b}
\end{equation}
where the exponent of $r$ could take values other than $2$ and was
restricted by $|b| \ge 2$. In the present case if one inserts the
$\phi (r)$ from \eqref{Misoln} into the second equation in
\eqref{g} (the first is automatically satisfied via the choice in
\eqref{solution1}) one obtains
\begin{equation}
\label{Misoln1}\left( \frac{\lambda ^{\prime
\prime}}{\lambda}- \frac{3}{2}\left(\frac{\lambda
^{\prime}}{\lambda }\right)^2 - \frac{\lambda  ^{\prime}}{r
\lambda } \right) ~ \rightarrow ~ \frac{4 - b^2}{2 r^2}
\end{equation}
which only vanishes for the choice $b = \pm 2$. Thus for $n>2$ the
$b = \pm 2$ solution of \cite{Mi} is selected. In \cite{Mi} the $b
\ge 2$ solutions gave a scale factor, $\phi (r)$, which monotonically increased as
one moved away from the brane. The $b \le 2$ solutions
gave a scale factor, $\phi (r)$, which decreased monotonically as one
moved away from the brane. For the solution in
\eqref{non-singular} the increasing $\phi (r)$ corresponds to
$a>1$ and the decreasing $\phi (r)$ to $a<1$. It was pointed out
in \cite{Mi} that both increasing and decreasing solutions give
localization of fields with spins ranging from $0$ to $2$.

Some of the parameters in \eqref{non-singular} can be fixed. The
width of the brane, $\epsilon$, is given by the inflection point
of $\phi (r)$, {\it i.e.} $\phi '' (r=\epsilon ) =0$. This
conditions gives $c=3 \epsilon ^2$. Next we require that $\lambda
(r=0) = 1$, {\it i.e.} that on the brane one has a 6-dimensional
Minkowski metric. This gives $\rho ^2 = \frac{3 \epsilon
^2}{2(a-1)}$. Putting these parameters back into
\eqref{non-singular} yields
\begin{equation}
\label{non-singular1} \phi (r) = \frac{3\epsilon^2 + a r^2}{3
\epsilon ^2 + r^2} ~~~~,~~~~ \lambda (r) = \frac{9 \epsilon ^4}{(3
\epsilon ^2 + r^2)^2}~,
\end{equation}
Now taking $\phi$ and $\lambda$ from the above equation and
inserting them into equations \eqref{Einstein-na} and
\eqref{Einstein-nb} we obtain expressions for the source functions
$F(r)$ and $K(r)$
\begin{eqnarray}
\label{FKa} \frac{F(r)}{M^{n+2}} = \Lambda &+& \frac{-6 \epsilon^2
n (n-3a+2)}{(3 \epsilon ^2 + a r^2)^2} + \frac{2 \left(3(n+2)+3
a^2 (n+2)-2 a (n^2+2n+6)\right) r^2}{(3 \epsilon ^2 + a r^2)^2} \nonumber \\
&+& \frac{-2 a n \left(-3+a(n+2) \right) r^4}{3 \epsilon
^2(3 \epsilon^2 + a r^2)^2} \\
\label{FKb} \frac{K(r)}{M^{n+2}} = \Lambda &+& \frac{-6 \epsilon^2
(n-1) (n-4a+2)}{(3 \epsilon ^2 + a r^2)^2} + \frac{4 \left(
2(n+2)+ 2 a^2 (n+2)-a (n^2+n+10) \right) r^2}{(3
\epsilon ^2 + a r^2)^2} \nonumber \\
&+& \frac{-2 a (n-1) \left( -4+a(n+2) \right) r^4}{3 \epsilon^2 (3
\epsilon ^2 + a r^2)^2}
\end{eqnarray}
For $n=2$ and using the relationship $\epsilon ^2 = \frac{40
M^4}{3 \Lambda}$ from \cite{GoSi} one can show that $F(r), K(r)$
reduce to the expressions in \cite{GoSi}. It is straightforward to
check that the last equation \eqref{Einstein-nc} is also satisfied
by $\phi(r), \lambda (r)$ from \eqref{non-singular1} and $K(r)$
from \eqref{FKb}. One can also check that the stress-energy tensor
defined by $F(r)$ and $K(r)$ above satisfy energy-momentum
conservation given by
\begin{equation}
\label{energy-con} \nabla ^A T_{AB} = \frac{1}{\sqrt{-^D g}}
\partial _A (\sqrt{-^D g}T^{AB}) + \Gamma ^B _{CD} T^{CD} = 0
\end{equation}
In terms of the source ansatz functions this yields
\begin{equation} \label{deltaT}
K^{\prime} + 4 \frac{\phi^\prime}{\phi} \left(K - F \right) = 0 ~.
\end{equation}
Using the $F(r)$ and $K(r)$ from \eqref{FKa} one finds that indeed
\eqref{deltaT} is satisfied. Thus taken together
\eqref{non-singular1} \eqref{FKa} \eqref{FKb} give a non-singular,
trapping solution to the $4+n$ dimensional Einstein equations
\eqref{Einstein-na} \eqref{Einstein-nb} \eqref{Einstein-nc}.

\section{Localization of Zero Modes}

As with the 6D solution of \cite{GoSi, Mi} the present solution
provides a gravitational trapping for fields of various spins: $0,
\frac{1}{2}, 1, 2$. The condition for a field to be trapped on the
brane is that the integral of the Lagrangian over the extra
coordinates is finite.
\begin{equation}
\label{trapping} S_s=\int d^D x \sqrt{-^D g} ~ L_s = \int d \Omega
_{n-1} \int dr \phi^4 \lambda ^{n/2} r^{n-1} \int d^4 x L_s
\end{equation}
where $L_s$ is the matter Lagrangian for fields with spin $s=0,
\frac{1}{2}, 1, 2$. We will outline the trapping of the zero modes
for each of these types of fields by applying the solution
\eqref{non-singular1} to the analysis given in \cite{Od1}. The
Lagrangians for the various spin fields will contribute different
factors of $\phi, \lambda$ and $r$ to the common factor of
$\sqrt{-^D g} = \phi^4 \lambda ^{n/2} r^{n-1}$ in
\eqref{trapping}.

For a scalar field the specific from of the action is
\begin{equation}
\label{scalar} S_0=-\frac{1}{2} \int d^D x \sqrt{-^D g}~ g^{MN}
\partial _M \Phi \partial _N \Phi
\end{equation}
The equations of motion from this scalar Lagrangian have a
zero-mode solution \cite{Od1} of the form $\Phi ( x^M ) = \phi
(x^\mu ) u_0$, with $u_0$ being the constant solution for the
extra dimensions, and $\phi (x^\mu)$ satisfies the Klein-Gordon
equation on the brane $\eta ^{\mu \nu}\partial _\mu \partial _\nu
\phi (x)=0$. Thus the scalar action becomes
\begin{equation}
\label{scalar1} S_0 = -\frac{\pi ^\frac{n}{2}}{\Gamma \left(
\frac{n}{2} \right)} u_0 ^2 \int _0 ^\infty dr \phi ^2 \lambda
^{n/2} r^{n-1} \int d^4x ~ \eta ^{\mu \nu} \partial _\mu \phi
\partial_\nu \phi
\end{equation}
The standard gamma function above comes from the integral over $d
\Omega _{n-1}$ namely $\int d \Omega _{n-1} =\frac{2 \pi
^\frac{n}{2}}{\Gamma \left( \frac{n}{2} \right)}$. The extra
factor of $\phi ^{-2}$ comes from $g^{\mu \mu}$ in the Lagrangian.
Inserting the solution into the radial integral we obtain
\begin{eqnarray}
\label{scalar2}\int _0 ^\infty dr \phi ^2 \lambda ^{n/2} r^{n-1}
&=& 3^n \epsilon ^{2n} \int _0 ^\infty r^{n-1} \frac{(3\epsilon ^2
+a r^2) ^2}{(3 \epsilon ^2 + r^2)^{n+2}} dr \nonumber \\
&=& \frac{3 ^{n/2} \epsilon ^n}{2^{n+1}}\left(a\sqrt{\pi}
\frac{\Gamma \left( 1+ \frac{n}{2} \right)}{\Gamma \left(
\frac{3+n}{2} \right)} + 2^n (1+a^2)\frac{\Gamma \left( 2+
\frac{n}{2} \right)\Gamma \left( \frac{n}{2} \right)}{\Gamma
(2+n)} \right)
\end{eqnarray}
which is finite.

The action for the spinor case is
\begin{equation}
\label{spinor} S_{1/2} = \int d^D x \sqrt{- ^D g} ~ {\bar \Psi} i
\Gamma ^M D_M \Psi .
\end{equation}
The curved spacetime gamma matrices ($\Gamma ^M$) are related to
the flat spacetime gamma matrices ($\gamma ^M$) via the inverse
vielbein $e ^M _{\bar M}$, $\Gamma ^M = e ^M _{\bar M}\gamma
^{\bar M}$. The vielbein is defined via $g_{MN} = e ^{\bar M} _M e
^{\bar N} _N \eta _{{\bar M} {\bar N}}$. $D_M \Psi = (\partial _M
+ \frac{1}{4} \omega ^{{\bar N} {\bar P}} _M \gamma _{{\bar N}
{\bar P}}) \Psi$ is the covariant derivative. The spin connection
$\omega ^{{\bar N} {\bar P}} _M$ is defined in terms of the
veilbein and its derivatives in the standard way. In \cite{Od1}
the zero mode for the $4+n$ dimensional Dirac equation was found
to be of the form $\Psi (x^A) = \psi (x^\mu) u(r) \chi (\theta _i
)$, where $\psi (x^\mu)$ satisfied the massless 4 dimensional
Dirac equation, $\gamma ^\mu
\partial_\mu \psi =0$, and $\chi (\theta _i )$ satisfied the
angular part of the higher dimensional Dirac equation. The
specific forms of $\psi$ and $\chi$ were not needed in the
analysis of the trapping of the fields. The $u(r)$ part of the
higher dimensional spinor was found \cite{Od1} to have the form
\begin{equation}
\label{spinor1} u(r)= C_{1/2} \phi ^{-2} \left( \lambda
^{\frac{1}{2}}
 r \right) ^{-\frac{n-1}{2}}
 \end{equation}
 where $C_{1/2}$ is an integration constant. In contrast
to the other fields, for the spinor field the extra space part of
the wavefunction is not simply a constant. Inserting all this
 into the spinor action in \eqref{spinor} yields
 \begin{eqnarray}
 \label{spinor2}
 S_{1/2} &=& \int _0 ^\infty dr \phi ^3 \lambda ^{\frac{n}{2}}
 r^{n-1} u^2 (r) \int d \Omega _{n-1} \chi ^2 (\theta _i) \int
 d^4x {\bar \psi} i \gamma ^\mu \partial _\mu \psi \nonumber \\
 &=&\int _0 ^\infty dr \phi ^{-1} \lambda ^{\frac{1}{2}}
 \int d \Omega _{n-1} \chi ^2 (\theta _i) \int
 d^4x {\bar \psi} i \gamma ^\mu \partial _\mu \psi
 \end{eqnarray}
 The extra factor of $\phi$ in the first expression comes from
 the veilbein of the higher dimensional gamma matrix, $\Gamma ^A$.
 The integral over $d \Omega _{n-1}$ is finite \cite{Od1} and the integral
 over $r$ is
 \begin{equation}
 \label{spinor3}
\int _0 ^\infty dr \phi ^{-1} \lambda ^{\frac{1}{2}} = \int _0
^\infty dr \frac{3 \epsilon ^2}{3\epsilon ^2+ a r^2} =
\frac{\sqrt{3} \epsilon \pi}{2 \sqrt{a}}
\end{equation}
which is also finite. Thus the integral over the extra coordinates
is finite and the fermions are localized.

For the vector, spin 1 case the action is given by
\begin{equation}
\label{vector} S_1 = -\frac{1}{4} \int d^D x \sqrt{- ^D g}~ g^{MN}
g^{RS} F_{MR} F_{NS} ~,
\end{equation}
where $F_{MN} = \partial _M A_N - \partial _N A_M$. There is a
zero mode solution \cite{Od1} of the form $A_M (x^N) = a_\mu
(x^\nu) u_0$ with $u_0$ the constant solution over the $n$ extra
space dimensions and $a_\mu$ satisfies the normal 4 dimensional
vacuum Maxwell equation, $\partial ^\mu (\partial _\mu a_\nu -
\partial _\nu a_\mu )= \partial ^\mu f_{\mu \nu}=0$. Inserting all
this along with the solution of \eqref{non-singular1} into
\eqref{vector} gives
\begin{equation}
\label{vector1} S_1 =-\frac{\pi ^\frac{n}{2}}{2 \Gamma\left(
\frac{n}{2} \right)} u_0 ^2 \int _0 ^\infty dr \lambda ^{n/2}
r^{n-1} \int d^4x \eta ^{\mu \nu} \eta^{\lambda \sigma} f_{\mu
\lambda} f_{\nu \sigma}
\end{equation}
The radial integral is evaluated as
\begin{equation}
\label{vector2}\int _0 ^\infty dr \lambda ^{n/2} r^{n-1}=\int _0
^\infty dr \frac{3^n \epsilon ^{2n} r^{n-1}}{(3\epsilon^2 +
r^2)^n}=\frac{3^{n/2} \epsilon ^n \sqrt{\pi}~ \Gamma
\left(\frac{n}{2} \right)}{2^n \Gamma \left(\frac{1+n}{2} \right)}
\end{equation}
Since the radial integral is finite the vector fields are also
localized on the brane by this solution.

The analysis for the spin-2 is similar to the spin-0 case and
results in the same radial integral (see \cite{GoSi}). Thus the
present solution also localizes the 4-dimensional spin-2 graviton
on the brane, since the integral \eqref{scalar2} is finite. 
As for the 6D solution, the present solution localizes spin-$0,
\frac{1}{2}, 1, 2$ onto the brane via gravity alone. In all cases
the integral over the extra coordinates is finite independent of
whether the 4D scale function, $\phi ^2$, is increasing ($a>1$) or
decreasing ($a<1$). Thus, as pointed out in \cite{Mi}, both types
of solutions localize the fields. This is connected with the fact
that the extra space scale function, $\lambda$, is independent of
$a$, and decreases to zero away from the brane in all cases.

\section{Conclusions}

The above results are in some sense expected since the scale
factors, $\phi (r) , \lambda (r)$, given in \eqref{non-singular1}
are of the same form as in the 6D case in \cite{GoSi}. However in
contrast to the general 6D solution found in \cite{Mi}, the
present higher dimensional solution selects out the $b=\pm 2$
exponent for the $r^{b}$ dependence of $\phi (r)$.

The behavior of the ansatz functions, $F(r)$ and $K(r)$, depends
on the various parameters, $n, \epsilon, \Lambda$ and $a$. In
particular $\Lambda$ and $a$ play the major role in determining
the values of $F(r)$ and $K(r)$ at $r=0$ and $r=\infty$. For
a large choice of parameters which lead to $\delta$-like behavior
for $F(r)$ and $K(r)$ ({\it i.e.} peaked near $r=0$ and decreasing
as $r \rightarrow \infty$). Let us now focus on the $r \rightarrow
\infty$ limit of $F(r)$ and $K(r)$. It is possible to have either
$F(r)$ or $K(r)$ or both go to zero for some choices of
parameters, $a$, $\Lambda$, $\epsilon$, and $n$. To see this we
first write the asymptotic values of $F(r)$ and $K(r)$ from
\eqref{FKa}
\begin{eqnarray}
\label{FKa1} F(r \rightarrow \infty) &=& \Lambda M^{n+2}+ \frac{ 2
n M^{n+2} \left( 3  - a (n+2) \right)}{3 a \epsilon ^2}
\nonumber \\
K(r \rightarrow \infty) &=& \Lambda M^{n+2}+  \frac{2 (n-1)
M^{n+2}\left( 4 - a (n+2) \right)}{3 a \epsilon ^2}
\end{eqnarray}
It is possible to make $F(\infty ) = 0$ by choosing
\begin{equation}
\label{Fa0} a=\frac{6 n}{4n +2 n^2 -3 \epsilon ^2 \Lambda}
\end{equation}
and it is possible to make $K(\infty ) = 0$ by choosing
\begin{equation}
\label{Ka0} a=\frac{8 (n-1)}{2(n -1)(n+2) -3 \epsilon ^2 \Lambda}
\end{equation}
If $\Lambda =0$ in \eqref{Fa0} and \eqref{Ka0} then $a=
\frac{3}{n+2}$ and $a=\frac{4}{n+2}$ respectively. Thus for
$\Lambda =0$ one can only have $F(r)$ or $K(r)$ asymptotically go
to zero with $a<1$ solutions. One must have $\Lambda >0$ if one
wants $F(r)$ or $K(r)$ asymptotically to go to zero {\it and} have
$a>1$. It is also possible to have both $F(r)$ and $K(r)$ go to
zero at $r \rightarrow \infty$. Setting \eqref{Fa0} equal to
\eqref{Ka0} and solving for $3 \epsilon ^2 \Lambda$ gives
\begin{equation}
\label{3el} 3 \epsilon ^2 \Lambda = \frac{2n(n+2)(n-1)}{n-4}
\end{equation}
Inserting this into either \eqref{Fa0} or \eqref{Ka0} gives $a=
\frac{4-n}{2+n}$ which yields $a<1$ for $n \ge 2$. For the 6D case
considered in \cite{GoSi} the focus was on $a>1$ solutions and
thus there it was not possible to have both $F(r)$ and $K(r)$
asymptotically go to zero. Thus there are two possibilities to
have both $F(\infty) =0$ and $K(\infty) =0$: $n=2$ and $n=3$ which
correspond to $a=\frac{1}{2}$ and $a=\frac{1}{5}$. For $n > 4$ and $a
< 0$ and the solution in \eqref{non-singular} becomes singular.
The $n=4$ case also has problems. First the scale function $\phi
(r)$ becomes singular at $r=\infty$. Also from \eqref{3el} one
finds that either $\epsilon$ or $\Lambda$ or both are infinite.
Finally, if one recomputes the source ansatz functions, $F(r),
K(r)$ in this case one finds that they always go to $\infty$ at
$r=\infty$. However an interesting result is that for $n=3$ one has
a non-singular brane solution with sources that approach zero away
from the brane. In \cite{GhRo} it was argued that for
local defects ({\it i.e.} solutions where the stress-energy was strictly 
zero outside the brane or decreased exponentially) it might
not be possible to construct non-singular solutions for $n>2$.
Our $n=3$ solution evades this restrictions since the stress-energy
goes to zero as a power law rather than exponentially or being strictly
zero outside the brane.

In this paper we have extended the non-singular, 6 dimensional
brane, world trapping solution of \cite{GoSi, Mi} to $n>2$ 
extra dimensions. This is more than a simple
mathematical exercise since, as pointed out in \cite{GhRo}, there
is a significant difference between $n=2$ and $n>2$ extra
dimensions. As with the 6D solution, the present solution provides
a universal, gravitational trapping mechanism for fields of
various spins from spin 0 to spin 2. Considering the magnitude of
the stress-energy ansatz functions necessary to realize this
trapping solution one finds physically reasonable properties for a
range of parameters $n, \Lambda, \epsilon$ and $a$. In addition
for $n=2$ and $n=3$ it is possible to have the entire
stress-energy tensor go to zero at $r=\infty$. For $n \ge 4$ one
can have either $F(r)$ or $K(r)$ go to zero, but not both. We have
not addressed the question of the stability of this brane solution
in the present paper.

\section{Acknowledgement} This work is supported by a 2004
Fulbright Scholars Grant. DS thanks Prof. Vitaly Melnikov for the
invitation to work at VNIIMS and the People's Friendship
University of Russia.

\end{document}